\documentclass[a4paper,fleqn,usenatbib]{mnras}

\usepackage[T1]{fontenc}
\usepackage{ae,aecompl}

\usepackage{graphicx}	
\usepackage{amsmath}	
\usepackage{amssymb}	
\usepackage{subfigure}
\usepackage{multirow}
\usepackage{xcolor}

\newcommand{\pcc}{\,{\rm cm}^{-3}}
\newcommand{\gcc}{\,{\rm g \, cm}^{-3}}
\newcommand{\pcs}{\,{\rm cm}^{-2}}

\newcommand{\kel}{\, {\rm K}}

\newcommand{\msun}{\, {\rm M}_\odot}
\newcommand{\nh}{n_{\rm H}}
\newcommand{\Nhcol}{N_{\rm H}}

\newcommand{\pc}{\, {\rm pc}}

\newcommand{\myr}{\, {\rm Myr}}
\newcommand{\kyr}{\, {\rm kyr}}
\newcommand{\ug}{\, {\rm \mu G}}

\newcommand{\kms}{\, {\rm km \, s^{-1}}}

\newcommand{\vcol}{v_{\rm col}}

\newcommand{\nthp}{N$_2$H$^+$}

\title[\nthp{} as a star formation tracer]{NEATH II: \nthp{} as a tracer of imminent star formation in quiescent high-density gas}

\author[Priestley et al.]{
  F. D. Priestley$^1$\thanks{Email: priestleyf@cardiff.ac.uk}, P. C. Clark$^1$, S. C. O. Glover$^2$, S. E. Ragan$^1$, O. Feh\'{e}r$^1$, L. R. Prole$^3$,
  \newauthor R. S. Klessen$^{2,4}$
\\
$^{1}$School of Physics and Astronomy, Cardiff University, Queen's Buildings, The Parade, Cardiff CF24 3AA, UK \\
$^{2}$Universit\"{a}t Heidelberg, Zentrum f\"{u}r Astronomie, Institut f\"{u}r Theoretische Astrophysik, Albert-Ueberle-Stra{\ss}e 2, D-69120 Heidelberg, Germany\\
$^{3}$Centre for Astrophysics and Space Science Maynooth, Department of Theoretical Physics, Maynooth University, W23 F2H6 Maynooth, Ireland \\
$^{4}$Universit\"{a}t Heidelberg, Interdisziplin\"{a}res Zentrum f\"{u}r Wissenschaftliches Rechnen, Im Neuenheimer Feld 205, D-69120 Heidelberg, Germany\\ 
}

\date{Accepted XXX. Received YYY; in original form ZZZ}

\pubyear{2023}

\begin{document}
\label{firstpage}
\pagerange{\pageref{firstpage}--\pageref{lastpage}}
\maketitle

\begin{abstract}

Star formation activity in molecular clouds is often found to be correlated with the amount of material above a column density threshold of $\sim 10^{22} \pcs$. Attempts to connect this column density threshold to a {\it volume} density above which star formation can occur are limited by the fact that the volume density of gas is difficult to reliably measure from observations. We post-process hydrodynamical simulations of molecular clouds with a time-dependent chemical network, {and investigate the connection between commonly-observed molecular species and star formation activity}. We find that many molecules widely assumed to {specifically trace the dense, star-forming component of molecular clouds} (e.g. HCN, HCO$^+$, CS) actually also exist in substantial quantities {in material only transiently enhanced in density, which} will eventually return to a more diffuse state without forming any stars. {By contrast, \nthp{} only} exists in detectable quantities above a volume density of $10^4 \pcc$, the point at which CO, which reacts destructively with \nthp, begins to deplete out of the gas phase onto grain surfaces. This density threshold for detectable quantities of \nthp{} corresponds very closely to the volume density at which gas becomes irreversibly gravitationally bound in the simulations: the material traced by \nthp{} never reverts to lower densities, and quiescent regions of molecular clouds with visible \nthp{} emission are destined to eventually form stars. The \nthp{} line intensity is likely to directly correlate with the star formation rate {averaged} over timescales of around a Myr.

\end{abstract}

\section{Introduction}

Stars form in molecular clouds, specifically in the densest regions (often referred to as cores or clumps; \citealt{bergin2007}) where self-gravity is able to overcome the various forces opposing collapse. \citet{lada2010} proposed that star formation {\it only} occurs in regions with K-band extinctions $A_{\rm K} \gtrsim 0.8 \, {\rm mag}$, based on the correlation between the mass of material above this threshold and the star formation rate (as traced by the number of protostars in the cloud). Studies using entirely independent methods \citep[e.g.][]{konyves2015} have found evidence of similar thresholds for star formation activity, corresponding to a visual extinction of $A_{\rm V} \sim 8 \, {\rm mag}$ or a column density of $\Nhcol \sim 10^{22} \pcs$, but the significance of this value and the reasons for its existence remain unclear.

One limitation of observational studies is that they are restricted to probing the line-of-sight column density, while the {\it volume} density of the gas is presumably more relevant to its evolution; the Jeans mass, for example, scales with local gas properties ($M_J \propto \rho^{-1/2} \, T^{3/2}$; \citealt{klessen2016}) rather than the column of material. \citet{lada2010} argue that their column density threshold corresponds to a volume density of $\nh \sim 10^4 \pcc$, but this relies on the assumption of a specific cloud geometry, and in reality one would expect a distribution of volume densities along the line of sight, rather than a single value. Establishing line-of-sight structure in molecular clouds is fraught with difficulties; see, for example, the ongoing debate about whether the Musca cloud is a filament or an edge-on sheet \citep{tritsis2022,kaminsky2023}.

A common approach to mitigate this issue is to assume that the high volume-density gas associated with star formation is specifically traced by molecular lines with high critical densities, most commonly the $J=1-0$ HCN transition. However, the emission in these lines is typically dominated by material with significantly lower density ($\sim 10^3 \pcc$; \citealt{pety2017,kauffmann2017,evans2020,jones2023,neumann2023}) than the assumed threshold for star formation, making them unsuitable for this purpose. The one exception appears to be the \nthp{} $J=1-0$ line, which is almost exclusively detected in regions with column densities above $10^{22} \pcs$ \citep{kauffmann2017,tafalla2021}, coinciding quite closely with the \citet{lada2010} column density threshold for star formation.

Another limitation is that observations can only capture a single moment of a molecular cloud's $\myr$-scale lifetime. While previous levels of star formation activity can be measured, such as by counting the number of protostellar objects as in \citet{lada2010}, the potential of a cloud for future star formation is not easily accessible. Many studies divide objects between those which are gravitationally bound, and so presumably destined to collapse, and unbound (and therefore transient) objects, but this distinction relies on a number of quantities which are either difficult or impossible to measure accurately \citep{enoch2008}, and neglects the possibility of continuing accretion of material \citep{rigby2021,anderson2021}.

In this paper, we post-process hydrodynamical simulations of molecular clouds with a time-dependent chemical network, and investigate molecular tracers of material which {\it will} form stars. We find that \nthp{} only exists in substantial quantities in gas which is irreversibly undergoing gravitational collapse, making it a unique signature of future star formation activity in molecular clouds.

\section{Method}

\begin{table*}
  \centering
  \caption{Simulation collision velocities, total durations, durations of star formation (time after formation of the first sink particle), final mass ($M_{\rm sink}$) and number ($N_{\rm sink}$) of sink particles, the final mass of gas with density above $10^3$ ($M_{\rm mid}$) and $10^4 \pcc$ ($M_{\rm dense}$), and the average star formation rate (defined as $M_{\rm sink}/t_{\rm SF}$).}
  \begin{tabular}{ccccccccc}
    \hline
    Model & $\vcol$/$\kms$ & $t_{\rm end}$/$\myr$ & $t_{\rm SF}$/$\myr$ & $M_{\rm sink}$/$\msun$ & $N_{\rm sink}$ & $M_{\rm mid}$/$\msun$ & $M_{\rm dense}$/$\msun$ & $\left< {\rm SFR} \right>$/$\msun \myr^{-1}$ \\
    \hline
    V3 & $3$ & $6.79$ & $1.22$ & $155$ & $86$ & $3341$ & $263$ & $127$ \\
    V7 & $7$ & $5.53$ & $1.70$ & $102$ & $105$ & $2986$ & $211$ & $60$ \\
    V10 & $10$ & $7.66$ & $1.13$ & $74$ & $35$ & $1610$ & $265$ & $65$ \\
    V14 & $14$ & $8.70$ & $-$ & $0$ & $0$ & $4$ & $0$ & $-$ \\
    \hline
  \end{tabular}
  \label{tab:models}
\end{table*}

\begin{table}
  \centering
  \caption{Elemental abundances used in the chemical modelling.}
  \begin{tabular}{ccccc}
    \hline
    Element & Abundance & & Element & Abundance \\
    \hline
    C & $1.4 \times 10^{-4}$ & & S & $1.2 \times 10^{-5}$ \\
    N & $7.6 \times 10^{-5}$ & & Si & $1.5 \times 10^{-7}$ \\
    O & $3.2 \times 10^{-4}$ & & Mg & $1.4 \times 10^{-7}$ \\
    \hline
  \end{tabular}
  \label{tab:abun}
\end{table}

\begin{figure*}
  \centering
  \includegraphics[width=\textwidth]{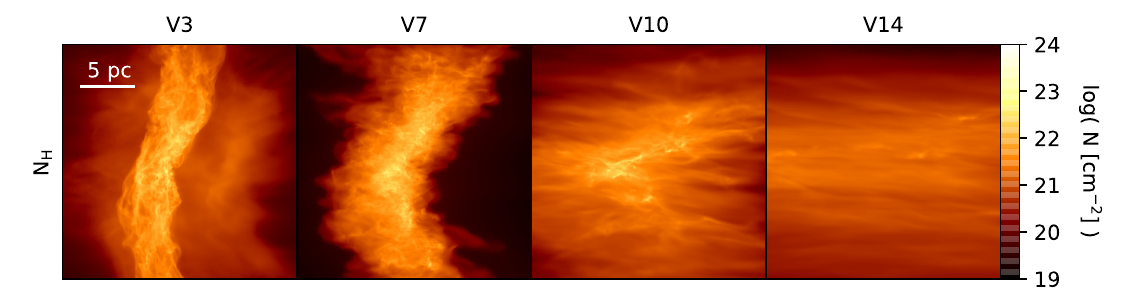}\\
  \includegraphics[width=\textwidth]{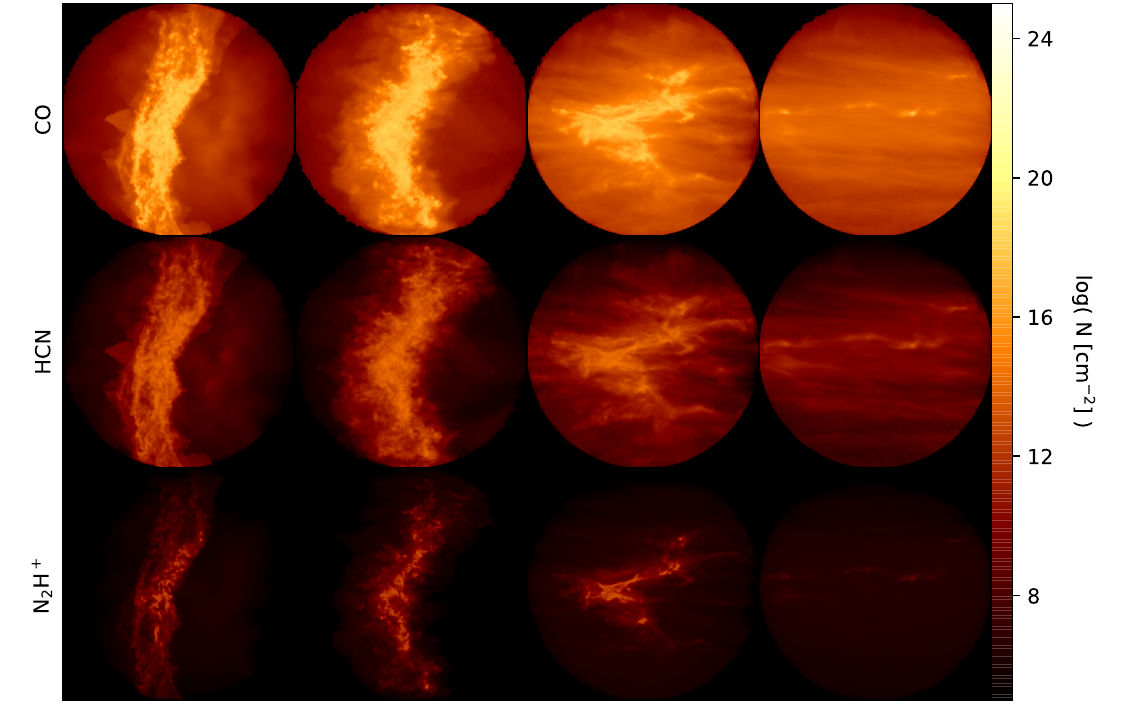}
  \caption{Column densities of hydrogen nuclei (top row), CO (second row), HCN (third row) and \nthp{} (bottom row) for the four simulations. The initial bulk velocities of the clouds are oriented along the horizontal axis. {Note that all post-processed tracer particles are located within a sphere of radius $16.2 \pc$, so the molecular column densities outside the projected boundaries of this sphere are set to zero.}}
  \label{fig:coldens}
\end{figure*}

\begin{figure}
  \centering
  \includegraphics[width=\columnwidth]{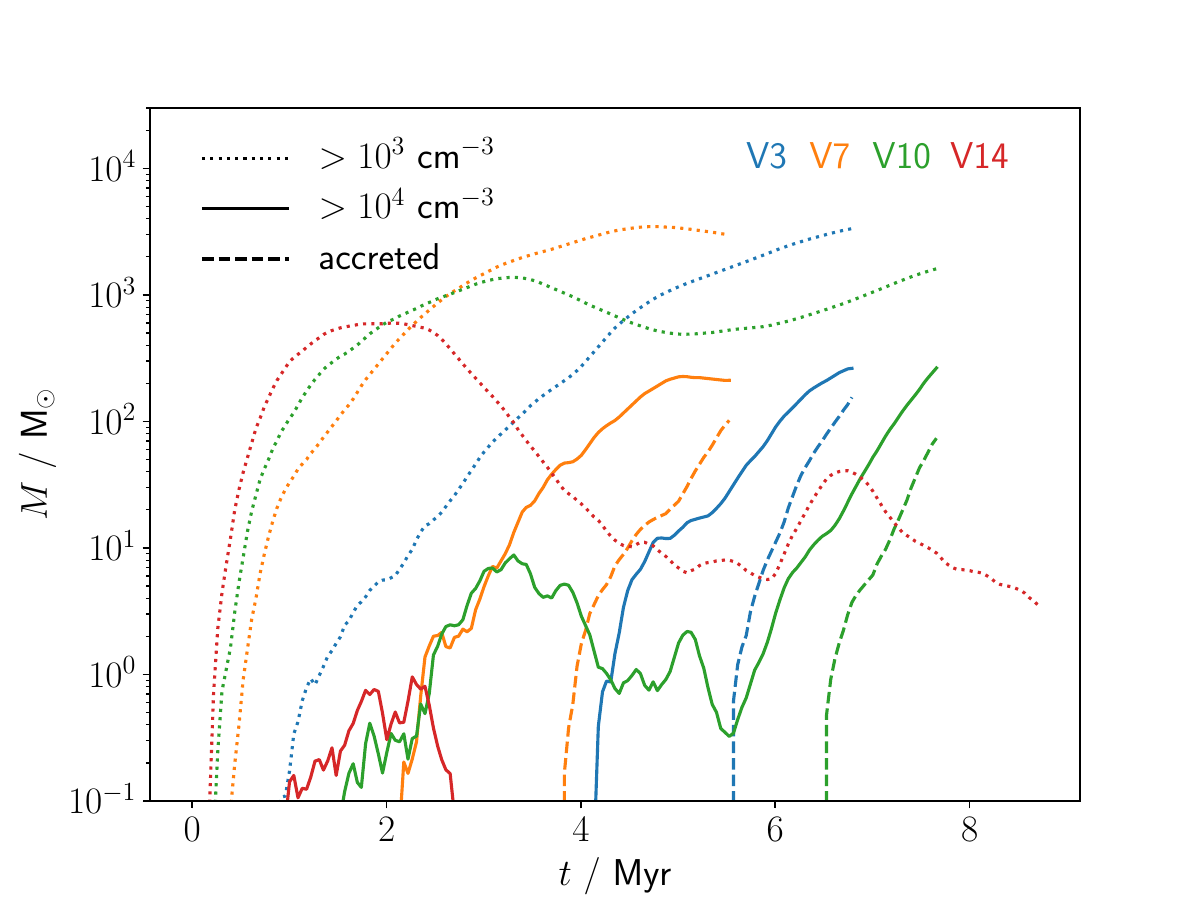}
  \caption{Time evolution of the mass of intermediate ($\nh > 10^3 \pcc$; dotted lines) and high ($\nh > 10^4 \pcc$; solid lines) density gas, and accreted sink mass (dashed lines), for the four simulations.}
  \label{fig:sfrate}
\end{figure}

\begin{figure}
  \centering
  \includegraphics[width=\columnwidth]{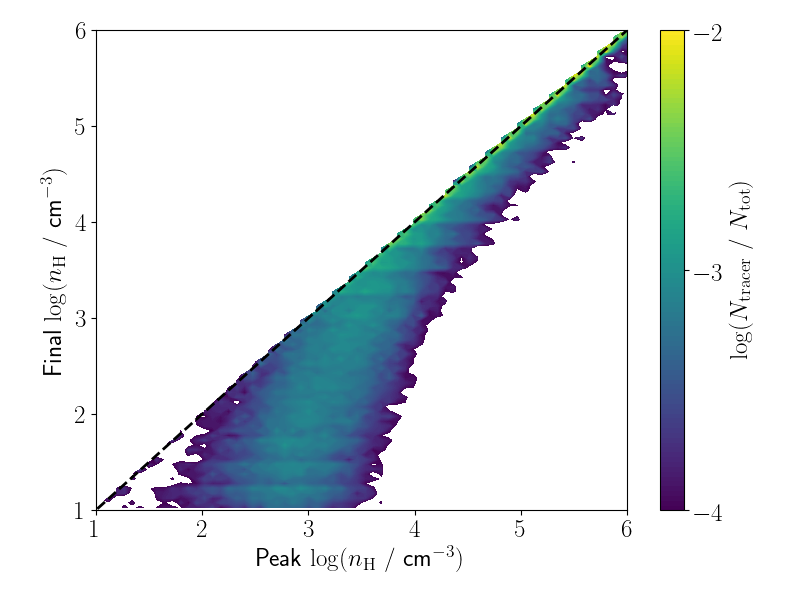}
  \caption{Distribution of tracer particles by peak density reached during the duration of the simulation, and the density at the simulation end, for the V7 model. The dashed black line shows the one-to-one relationship for tracer particles which only reach their peak density at the end of the simulation.}
  \label{fig:peaknumv7}
\end{figure}

\begin{figure*}
  \centering
  \includegraphics[width=0.3\textwidth]{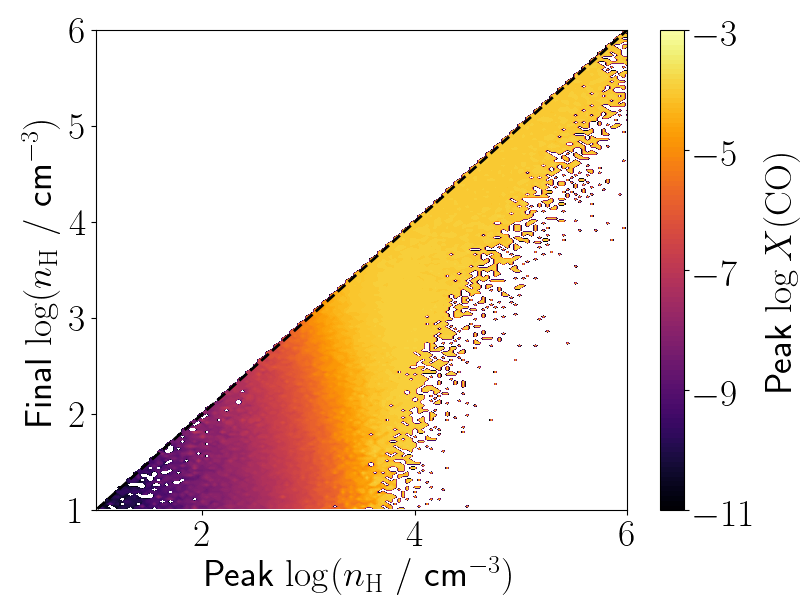}\quad
  \includegraphics[width=0.3\textwidth]{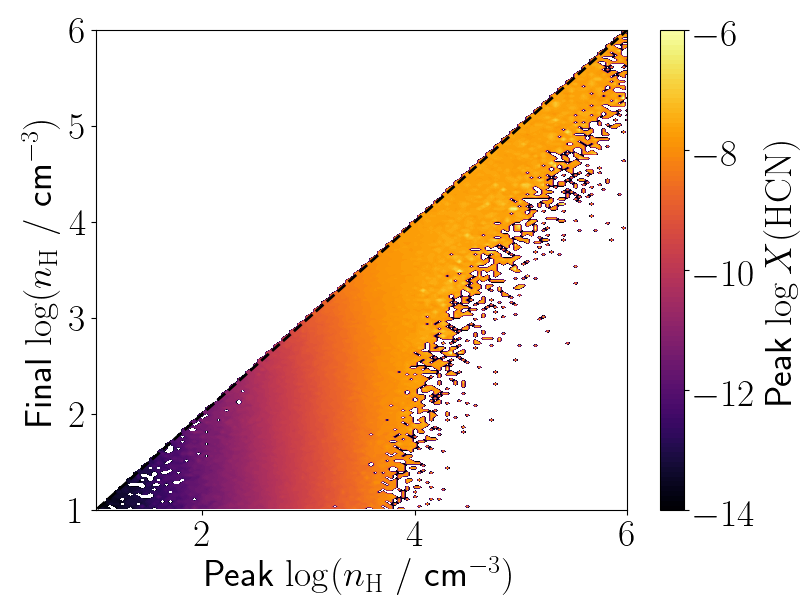}\quad
  \includegraphics[width=0.3\textwidth]{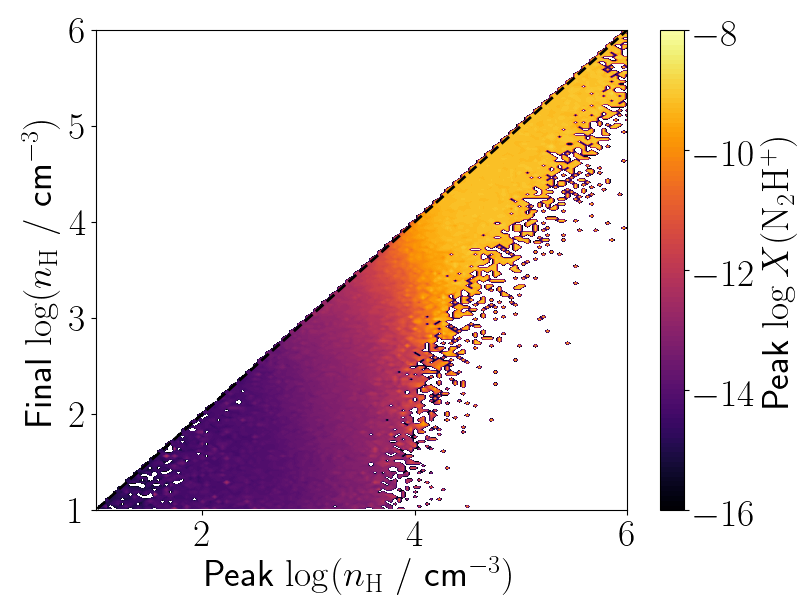}\quad
  \caption{Average peak gas-phase abundances of CO (left), HCN (centre) and \nthp{} (right), as a function of the peak density reached during the duration of the simulation, and the density at the simulation end, for the V7 model. The dashed black lines show the one-to-one relationship for tracer particles which only reach their peak density at the end of the simulation.}
  \label{fig:peakmolv7}
\end{figure*}

\begin{figure*}
  \centering
  \includegraphics[width=0.3\textwidth]{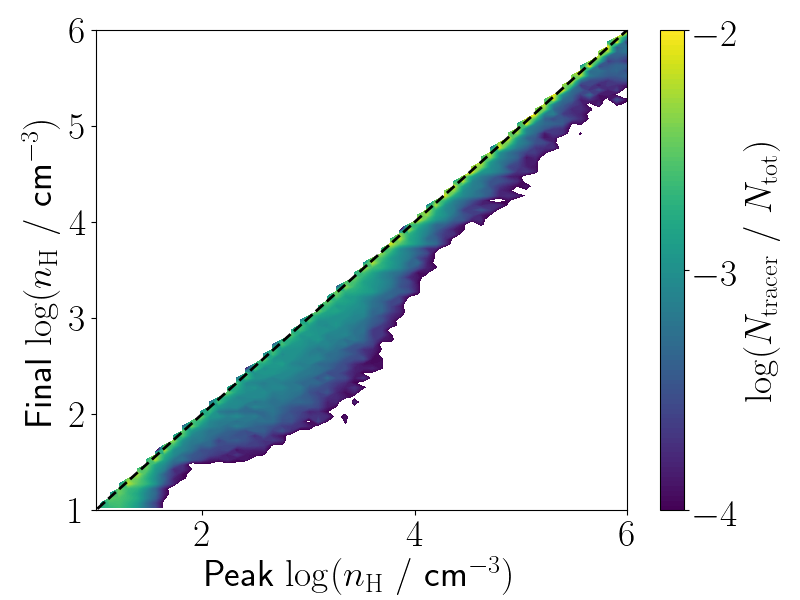}\quad
  \includegraphics[width=0.3\textwidth]{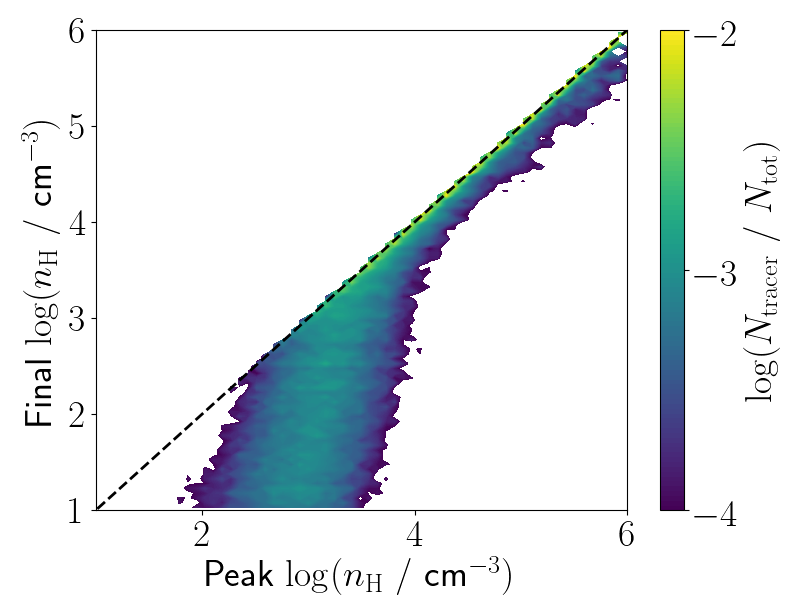}\quad
  \includegraphics[width=0.3\textwidth]{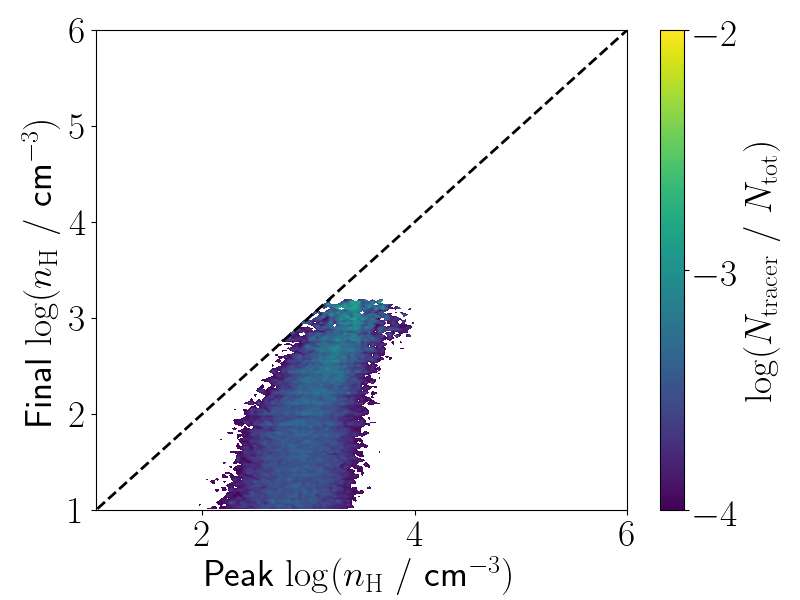}\quad
  \caption{Distribution of tracer particles by peak density reached during the duration of the simulation, and the density at the simulation end, for the V3 (left), V10 (centre) and V14 (right) models. The dashed black lines show the one-to-one relationship for tracer particles which only reach their peak density at the end of the simulation.}
  \label{fig:peaknumrest}
\end{figure*}

\begin{figure*}
  \centering
  \includegraphics[width=0.3\textwidth]{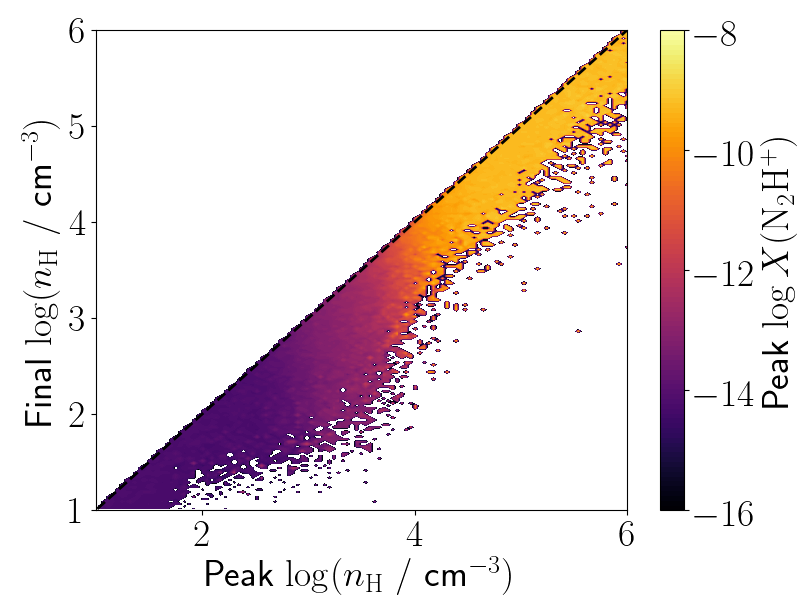}\quad
  \includegraphics[width=0.3\textwidth]{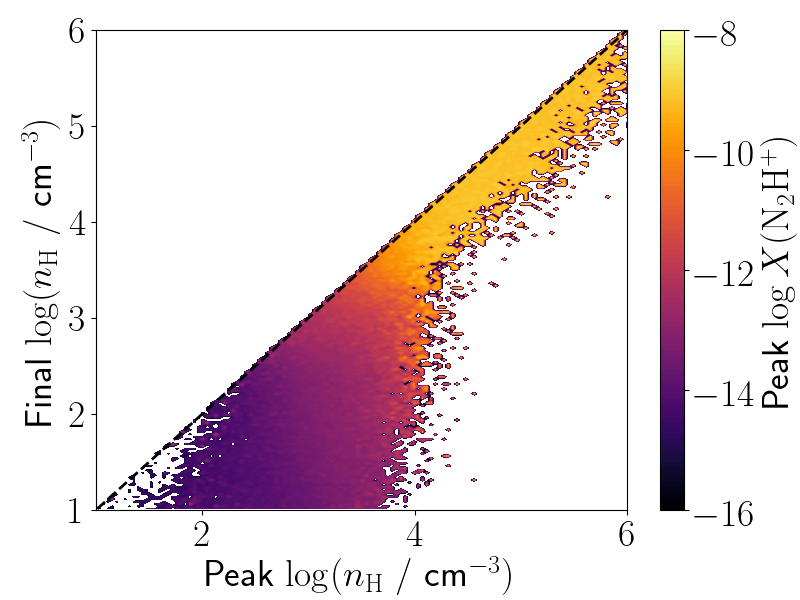}\quad
  \includegraphics[width=0.3\textwidth]{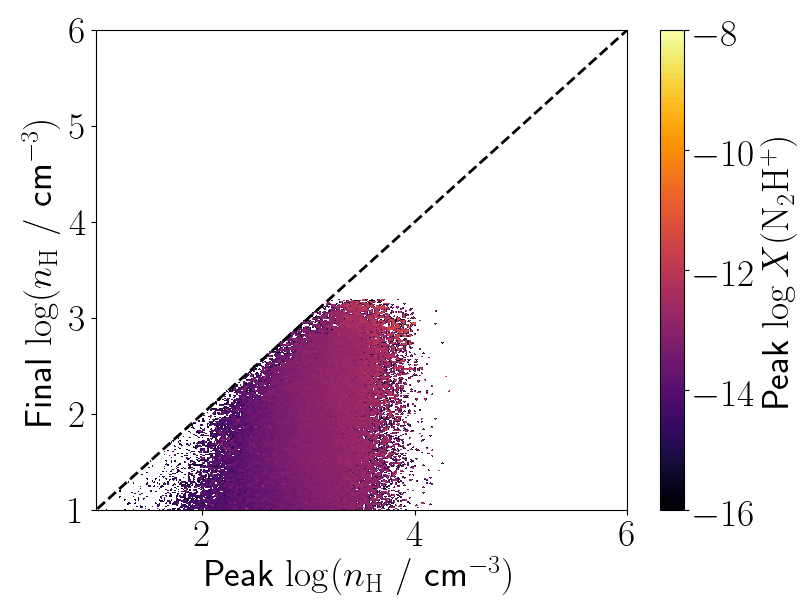}\quad
  \caption{Average peak gas-phase abundances of \nthp{} as a function of the peak density reached during the duration of the simulation, and the density at the simulation end, for the V3 (left), V10 (centre) and V14 (right) models. The dashed black lines show the one-to-one relationship for tracer particles which only reach their peak density at the end of the simulation.}
  \label{fig:peakmolrest}
\end{figure*}

\begin{figure*}
  \centering
  \includegraphics[width=0.3\textwidth]{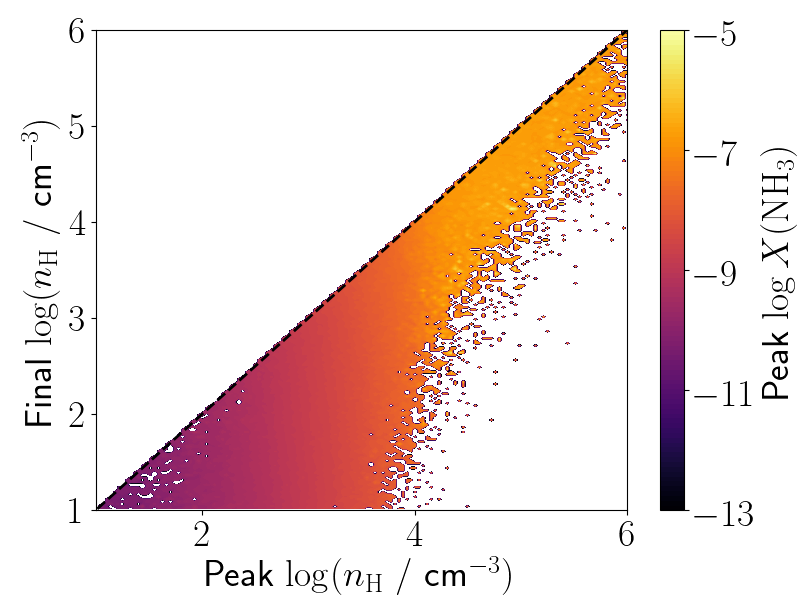}\quad
  \includegraphics[width=0.3\textwidth]{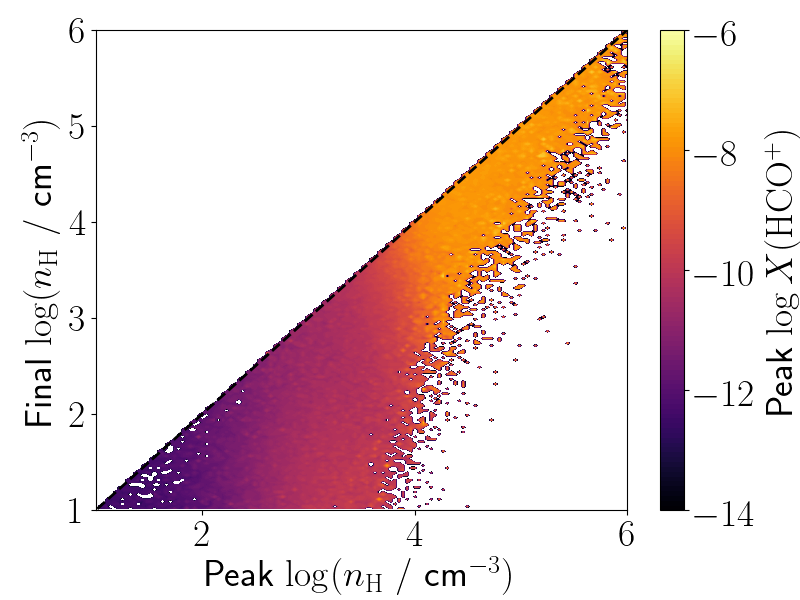}\quad
  \includegraphics[width=0.3\textwidth]{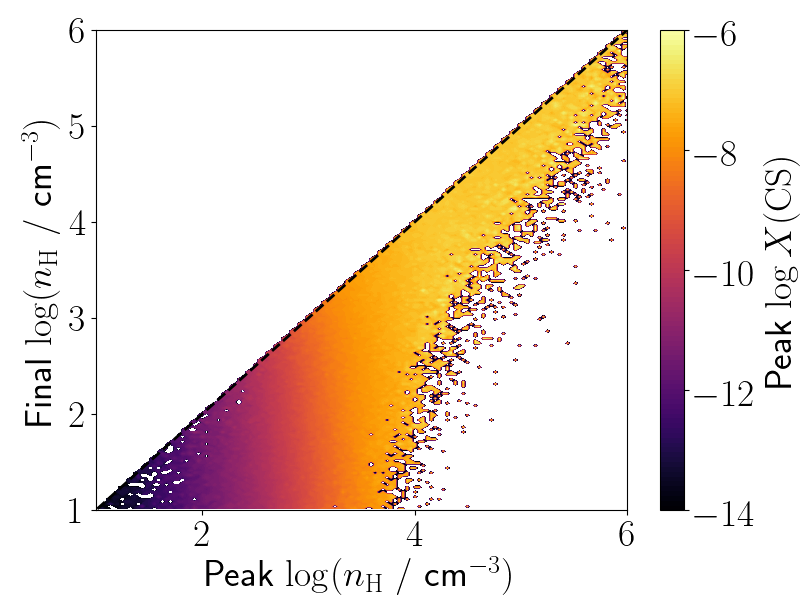}\quad
  \caption{Average peak gas-phase abundances of NH$_3$ (left), HCO$^+$ (centre) and CS (right), as a function of the peak density reached during the duration of the simulation, and the density at the simulation end, for the V7 model. The dashed black lines show the one-to-one relationship for tracer particles which only reach their peak density at the end of the simulation.}
  \label{fig:peakmolv7_2}
\end{figure*}

\begin{figure*}
  \centering
  \includegraphics[width=0.3\textwidth]{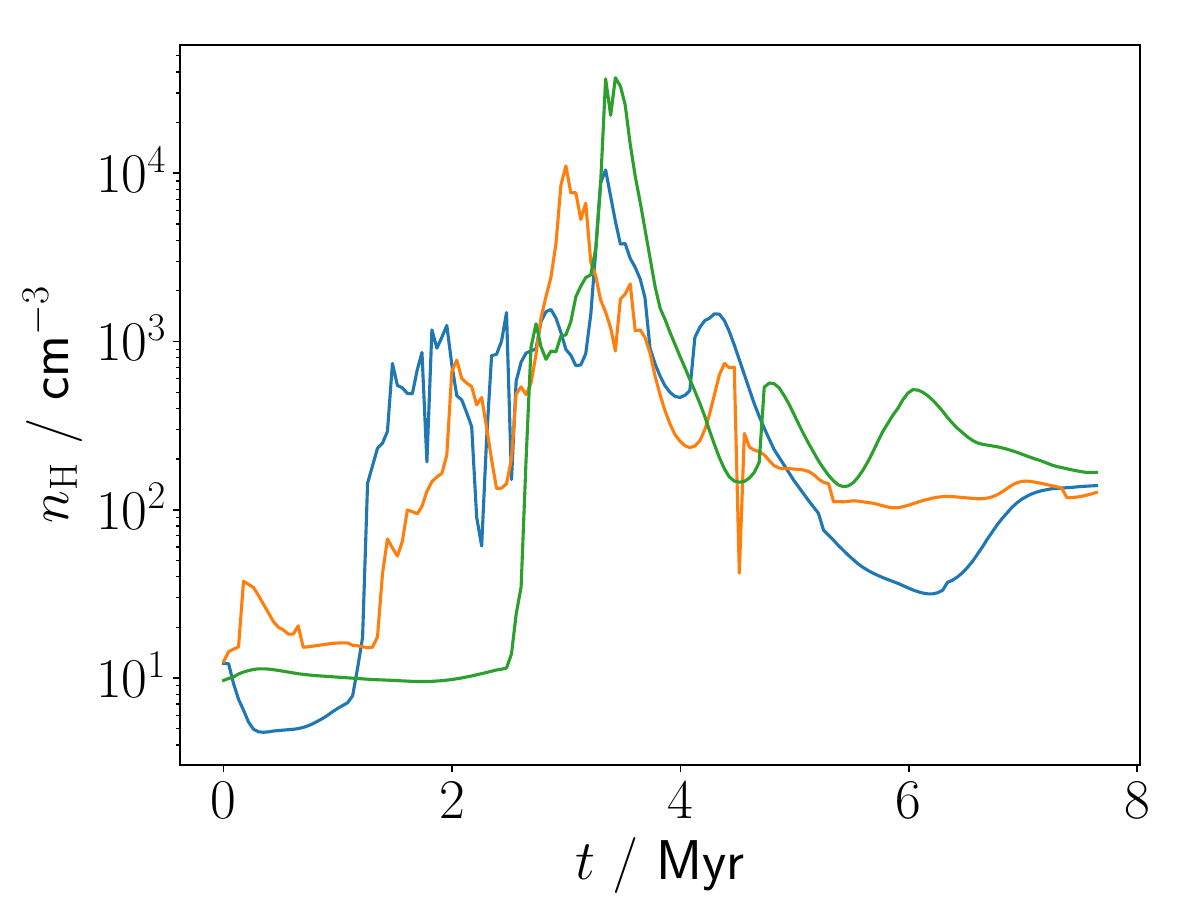}\quad
  \includegraphics[width=0.3\textwidth]{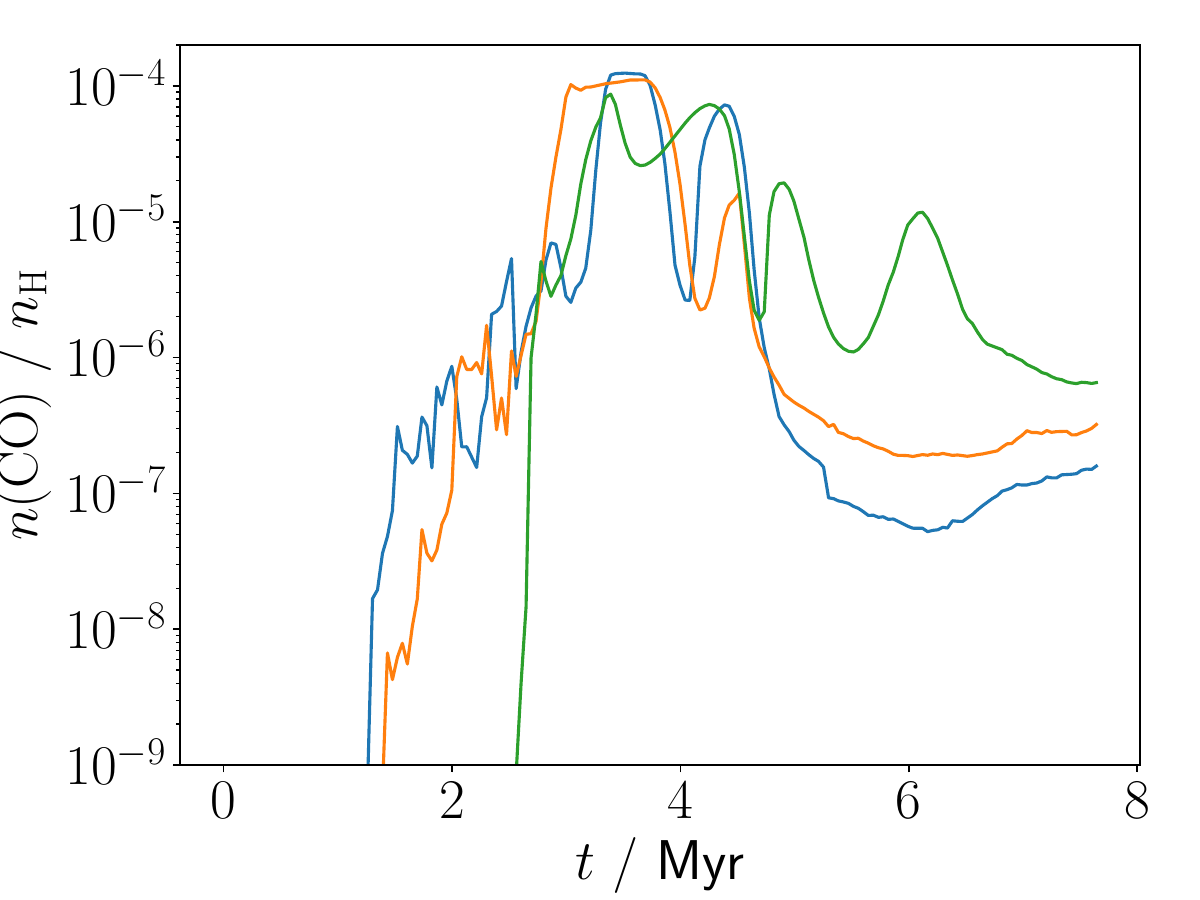}\quad
  \includegraphics[width=0.3\textwidth]{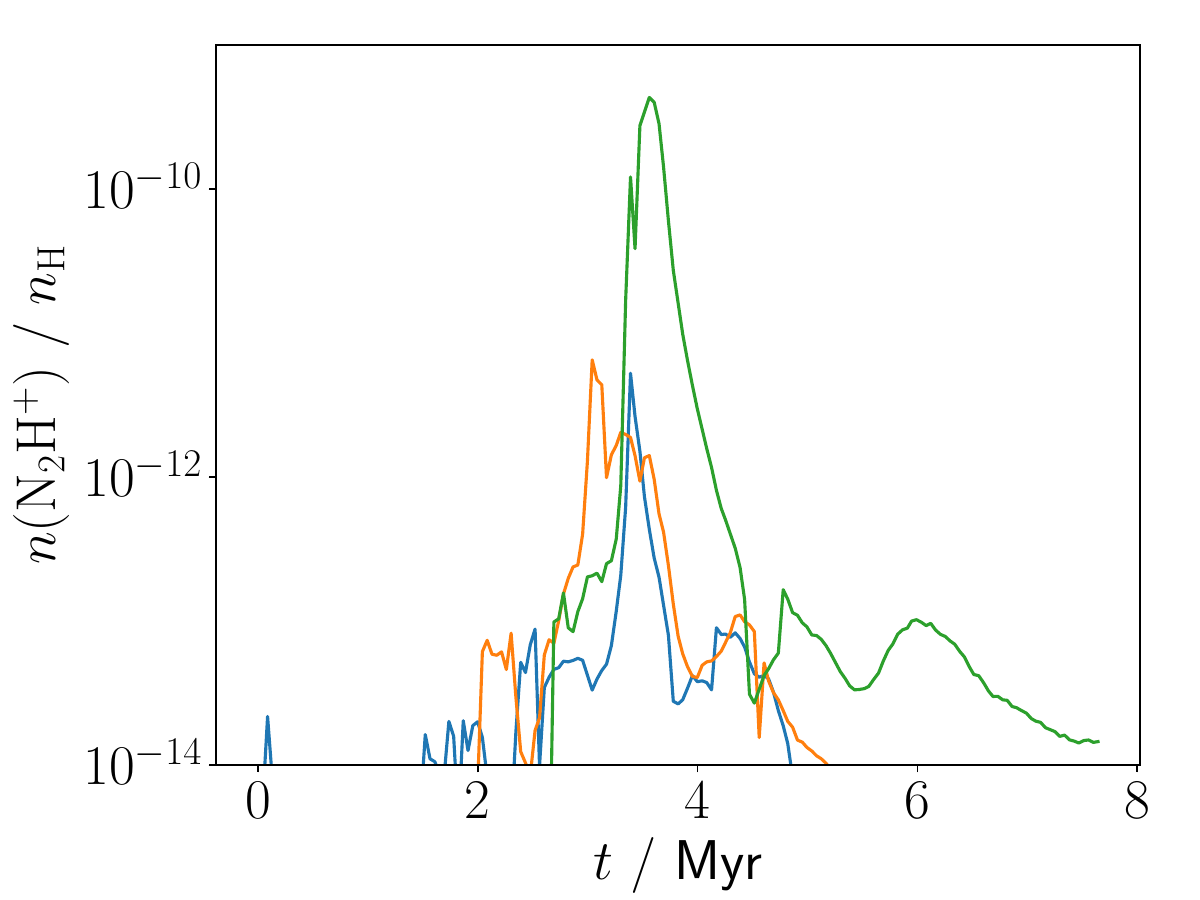}\quad
  \caption{Evolution of the gas density (left) and the CO (centre) and \nthp{} (right) abundances, for three tracer particles with peak density $> 10^4 \pcc$ and final density $< 10^3 \pcc$ from the V10 model.}
  \label{fig:trans}
\end{figure*}

\begin{figure}
  \centering
  \includegraphics[width=\columnwidth]{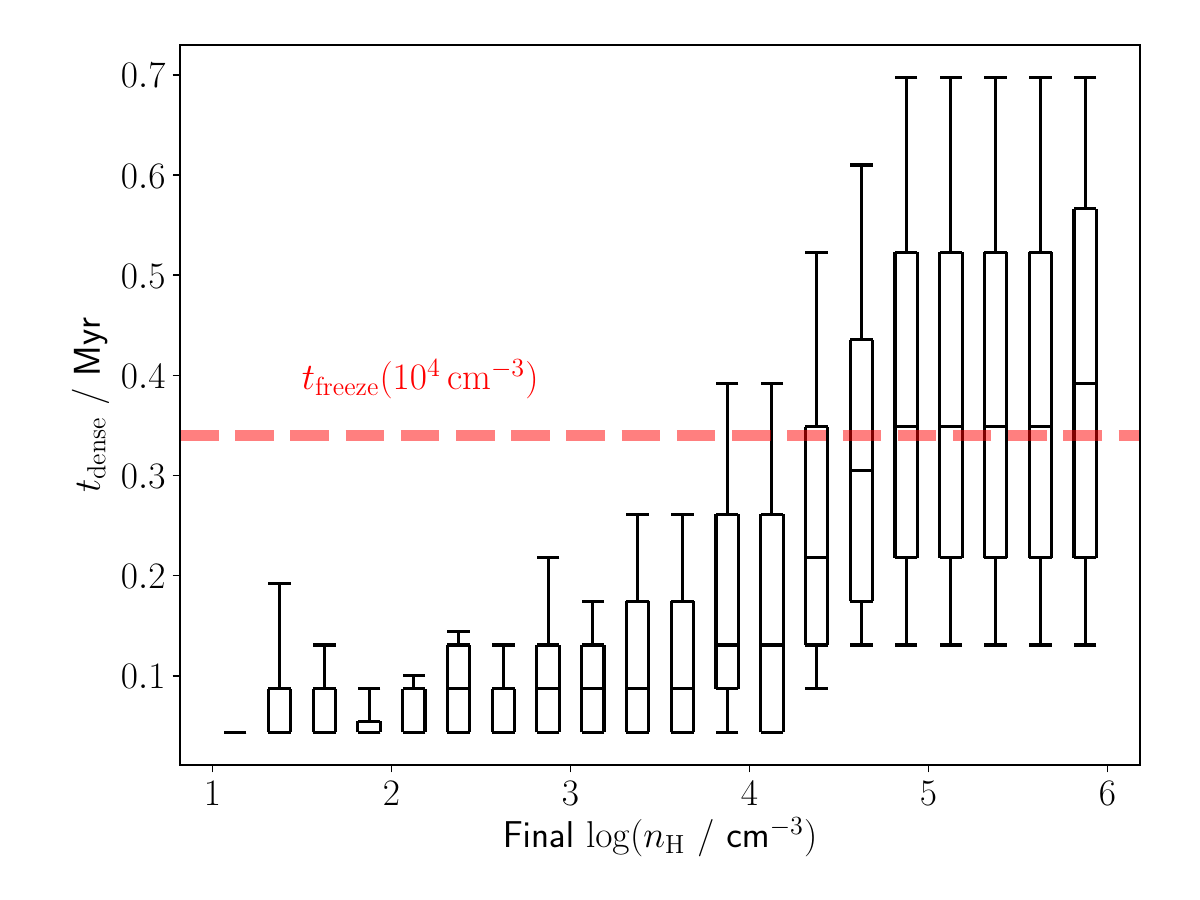}
  \caption{Distribution of time spent above $10^4 \pcc$ ($t_{\rm dense}$) versus final density, for all tracer particles with a peak density above $10^4 \pcc$ in the V10 model. Boxes show the median and 25th/75th percentiles, whiskers the 10th and 90th percentiles. The dashed red line shows the freeze-out timescale at a density of $10^4 \pcc$ ($0.34 \myr$).}
  \label{fig:tfreeze}
\end{figure}

We perform simulations of colliding, initially-atomic clouds using {\sc arepo} \citep{springel2010}, a moving-mesh magnetohydrodynamics (MHD) code. The simulations include an extensive suite of processes affecting the thermodynamics of the gas and dust \citep{glover2007,glover2012}, including an on-the-fly chemical network (a modified version of the \citealt{gong2017} network, as described in \citealt{hunter2023}) for the key atomic and molecular coolants, with self-shielding from the external ultraviolet (UV) radiation field calculated self-consistently \citep{clark2012}.

We initialise two spherical clouds of mass $10^4 \msun$ and radius $19 \pc$, for an initial volume density of $10 \pcc$ and a gas (dust) temperature of $300 \kel$ ($15 \kel$), displaced along the $x$-axis so that the clouds are initially just touching. The clouds have a virialised turbulent velocity field with a root-mean-square value of $0.95 \kms$, and a bulk motion along the $x$-axis $v_x = \pm \vcol$ for the cloud displaced in the $\mp x$ direction; we investigate four values of $\vcol$ between $3$ and $14 \kms$, giving net collision velocities $2 \vcol$ between $6$ and $28 \kms$. Model properties are listed in Table \ref{tab:models}. A $3 \ug$ magnetic field is included parallel to the collision axis. Sink particles are introduced following \citet{tress2020} (see also \citealt{prole2022}), with a threshold density of $2 \times 10^{-16} \gcc$ and formation radius of $9 \times 10^{-4} \pc$.

We use Monte Carlo tracer particles \citep{genel2013} to record the evolution of the physical properties of individual parcels of gas. We then post-process the time-dependent chemical evolution of the tracers with the NEATH framework\footnote{https://fpriestley.github.io/neath/} \citep{priestley2023c}, using a modified version of the {\sc uclchem} code \citep{holdship2017} with the UMIST12 reaction network \citep{mcelroy2013}. We assume elemental abundances from \citet{sembach2000} with the refractory elements depleted by an additional factor of 100 (as is appropriate for the dense interstellar medium; \citealt{lee1998,jenkins2009}), listed in Table \ref{tab:abun}. The external UV field is set to $1.7$ Habing units \citep{habing1968}, and the cosmic ray ionisation rate to $10^{-16} \, {\rm s^{-1}}$. These chemical parameters are shared with the underlying {\sc arepo} simulations.

We select $10^5$ tracer particles from each simulation to model chemically, chosen from particles with a final position within $16.2 \pc$ of the simulation centre; this region includes virtually all of the dense gas in the simulations. Particles are chosen to evenly sample final densities in the range $10-10^6 \pcc$, with the exception of the V14 simulation, where the upper boundary is the maximum density present at the simulation end point ($2000 \pcc$). The limit of $10^6 \pcc$ is motivated by the $44 \kyr$ time resolution of the simulations (i.e. the interval between updates of tracer properties), {which is insufficient to fully resolve the freefall time of gas at higher densities. This leads to the rapidly-evolving physical properties in collapsing regions not being captured by the chemical modelling, making the resulting molecular abundances unreliable. A time resolution of $44 \kyr$ produces abundances which are converged up to gas densities of $10^5 \pcc$, and accurate to within a factor of two up to $10^6 \pcc$ \citep{priestley2023c}, which is more than sufficient for our purposes.}

\section{Results}

\subsection{Physical properties}

Table \ref{tab:models} lists the mass and number of sinks formed in each of the four simulations, and the mass of gas above volume density thresholds of $10^3$ and $10^4 \pcc$ at the final time $t_{\rm end}$. We choose $t_{\rm end}$ so that the V3, V7 and V10 simulations have all formed a similar quantity of dense ($\nh > 10^4 \pcc$) gas; the V14 simulation is evolved to the point where it appears to be dispersing, as it never forms any substantial amount of dense gas.

Figure \ref{fig:coldens} shows the simulation column densities, viewed perpendicular to the collision axis at $t_{\rm end}$. The V3 and V7 models have both formed a compressed layer at the interface between the two clouds, which contains the vast majority of the dense material. In the V10 simulation, the higher kinetic energy of the initial cloud motions has disrupted this layer, and the dense material is more spread out along the collision axis. In the V14 simulation, the collision velocity is so high that no layer forms, and there is very little visible dense substructure. Some of the overdense regions that do form may eventually collapse under {their own} self-gravity (as in \citealt{jones2023}), but there is no sign of this happening at the point we terminate the simulation.

Figure \ref{fig:sfrate} shows the time evolution of the masses of intermediate density ($M_{\rm mid}$; $\nh > 10^3 \pcc$) and genuinely dense gas ($M_{\rm dense}$, defined using a volume density threshold of $10^4 \pcc$ as suggested by \citealt{lada2010}), and the total mass accreted onto sink particles ($M_{\rm sink}$). Intermediate density gas forms earlier for faster collision velocities, but the two higher velocity models subsequently show a decline in the mass of this material, which for the V14 model appears to be irreversible. The onset of dense gas and sink formation has a more complex relationship with $\vcol$; model V7 forms dense gas and sinks before V3, but has a slower rate of star formation, as indicated by the lower average values\footnote{We note that the star formation rates in our simulations are comparable to those estimated observationally by \citet{lada2010}.} in Table \ref{tab:models}. Model V10 first forms dense gas at a comparable time to V7, but this gas then disperses before a second round of dense gas formation several $\myr$ later, this time accompanied by sink formation. Model V14 never forms any significant quantity of dense gas, nor any sinks, over the runtime of the simulation. Star formation activity (as represented by sink particles) {only seems to occur after} the formation of a substantial ($>10 \msun$) mass of dense gas. The mass of stars formed approximately tracks the mass of {this} dense gas with a delay of $\sim 1-2 \myr$.

\subsection{Chemical properties}

Figure \ref{fig:coldens} shows the column densities of CO, which is typically thought to trace gas with densities above $\sim 100 \pcc$ \citep[e.g.][]{clark2019}, and HCN and \nthp, both commonly used as tracers of denser gas. Regions of significant CO column density exist above a threshold total column of $\sim 10^{21} \pcs$ \citep{priestley2023c}, but {much of the substructure in regions above this threshold is not visible in the CO column}; there is essentially no distinction between moderate- and high-density sightlines. HCN fares slightly better, but substantial columns are still present in fairly low-density sightlines, {in agreement with its widespread detection in regions of molecular clouds with $A_{\rm V} < 8 \, {\rm mag}$ \citep[e.g.][]{evans2020}.}

As {found} by previous observational studies \citep{kauffmann2017,tafalla2021}, \nthp{} {only exists in significant quantities} for $\Nhcol \gtrsim 10^{22} \pcs$ \citep{priestley2023c}. This is apparent in Figure \ref{fig:coldens}, where the morphology of the clouds seen in \nthp{} is very different to that in CO or HCN, with only the highest-density substructures being visible. The V14 model, which forms no sink particles, also has no substantial \nthp{} column density along any sightline. This suggests a possible role for the molecule as a direct tracer of material with the potential to form stars.

Figure \ref{fig:peaknumv7} shows the distribution of tracer particles\footnote{As the tracer particles do not correspond to physical gas parcels, this is not a direct measure of the mass distribution, but is a reasonable proxy quantity.} in the post-processed sample, according to the peak density reached at any point during their evolution, and the final density at $t_{\rm end}$, for the V7 model {(the intermediate case out of the three models which form sink particles)}. Tracer particles which fall below the one-to-one relationship (represented by the dashed black line) were, at some point, at a higher density than their final value; they have undergone a transient enhancement in density, which has not resulted in runaway gravitational collapse.

Up to a peak density of about $10^4 \pcc$, there is a significant population of tracers below the one-to-one line. These tracers reached relatively high densities {due to compression by supersonic gas motions}, but subsequently returned to much lower values\footnote{{This re-expansion is also due to the turbulent gas motions, as the magnetic field strength is - by design - too weak to prevent the collapse of gravitationally-bound structures.}}, extending down to the initial cloud density of $10 \pcc$. By contrast, this behaviour is almost non-existent above a peak density of $10^4 \pcc$. Material which reaches this density never becomes {strongly rarefied in its subsequent evolution, and is effectively destined to form (or accrete onto) a sink particle.} This corresponds very well with the volume density threshold proposed by \citet{lada2010}, and appears to be a natural outcome of the combined hydrodynamical and thermal evolution of molecular gas, {although the exact reasons for the existence and value of this threshold are currently unclear}.

Figure \ref{fig:peakmolv7} shows how the peak molecular abundances of CO, HCN and \nthp{} vary with the peak and final densities reached by the tracer particles, again for the V7 model. In addition to their {limited ability to discern high-density substructures}, CO and HCN both reach {high} peak abundances ({relative to their maximum gas-phase abundance at any density}) in material which reaches densities of a few thousand $\pcc$, but subsequently returns to $\sim 10 \pcc$ and thus plays no role in star formation. By contrast, \nthp{} has an abundance of $\lesssim 10^{-12}$ in this material, which we demonstrate in Section \ref{sec:radex} would be observationally challenging to detect. Significant quantities of this molecule are only produced in gas which reaches, and then stays above, the $10^4 \pcc$ density threshold.

Figures \ref{fig:peaknumrest} and \ref{fig:peakmolrest} show that this behaviour - for both the physical and chemical evolution - is independent of the simulation collision velocity. For all models, we find substantial quantities of gas which reaches a density above $10^3 \pcc$ and subsequently returns to much lower values, but {this behaviour is vanishingly rare} for material which exceeds a volume density of $10^4 \pcc$. This volume density threshold {for eventual gravitational collapse} corresponds very closely to the density at which significant quantities of \nthp{} are produced. We therefore suggest that detectable \nthp{} emission in a molecular cloud directly traces material which is destined to eventually form stars, rather than transiently-enhanced, unbound material, which may eventually disperse.

This behaviour is not shared by any other molecular species we investigate. Figure \ref{fig:peakmolv7_2} shows the behaviour of NH$_3$, HCO$^+$ and CS for the V7 model, {all of which are often {assumed to trace high-density, star-forming material}}. While NH$_3$ and HCO$^+$ reach their peak abundances only in material which exceeds the $10^4 \pcc$ threshold for {subsequent} collapse, both molecules also exist in appreciable quantities below this boundary. CS behaves more like HCN and CO, and is even more likely to be readily detectable in material which is only transiently enhanced in density. We address the detectability issue quantitatively in Section \ref{sec:radex} below.

\subsection{Transient high-density gas}
\label{sec:freeze}

{The two faster collisions, models V10 and V14, do appear to produce some dense gas which is not directly associated with star formation, reflected in the temporary peaks in dense gas mass between $\sim 2-3 \myr$ in Figure \ref{fig:sfrate}. Figure \ref{fig:trans} shows the time evolution of three tracer particles from the V10 model, which have peak densities above $10^4 \pcc$ but final densities significantly below this. The enhancement to high density is extremely brief ($\ll \myr$) in all cases, and the tracer particles quickly return to lower densities of $\sim 100 \pcc$.}

{The increase in density initially results in a rapid increase in the abundances of both CO and \nthp. However, once the CO abundance reaches $\sim 10^{-5}$, its ability to efficiently destroy \nthp{} terminates the rise in the abundance of the latter species, and two of the three tracer particles never exceed an \nthp{} abundance of $10^{-11}$. One tracer particle does reach a high enough density to begin depleting CO from the gas phase, allowing a peak \nthp{} abundance of $10^{-10}$ which we show in Section \ref{sec:radex} may be observationally relevant, but this only lasts for the brief period of time when the gas is at high density. Once the density begins to decline again, CO is rapidly desorbed back off grain surfaces, {as the rates of the main desorption processes (via cosmic rays and UV photons) decline more slowly (proportional to $\nh$) than freeze-out (proportional to $\nh^2$). This causes} a steep reduction in the abundance of \nthp.}

{Forming and maintaining a high \nthp{} abundance requires depletion of CO, which itself requires high densities. The rate of depletion onto grain surfaces is given by}
\begin{equation}
  \dot{n}_{\rm freeze} = 4.57 \times 10^4 \, \left< d_g a^2 \right> \, \left(T / m \right)^{0.5} \nh \, n \; \; {\rm cm^{-3} \, s^{-1}},
\end{equation}
{where $n$ is the gas-phase abundance of the molecule, $\pi \left< d_g a^2 \right>$ is the grain surface area per hydrogen nucleus, $T$ is the gas temperature, $m$ is the atomic mass of the molecule, and we have assumed a sticking coefficient of unity \citep{rawlings1992}. Taking $\nh = 10^4 \pcc$, $T = 20 \kel$, $m = 28$ for CO, and $\left< d_g a^2 \right> = 2.2 \times 10^{-22} \, {\rm cm^2}$ from \citet{rawlings1992}, the freeze-out timescale is $t_{\rm freeze} = n / \dot{n}_{\rm freeze} = 0.34 \myr$.}

{In Figure \ref{fig:tfreeze}, we compare this value with the amount of time spent by tracer particles above a density of $10^4 \pcc$, $t_{\rm dense}$ (CO freeze-out is inefficient at lower densities; \citealt{priestley2023c}), for the V10 model (we exclude tracer particles for which $t_{\rm dense} = 0$). Virtually all tracer particles with a final density below $10^4 \pcc$ have $t_{\rm dense} < t_{\rm freeze}$; material which reaches high densities only transiently does not spend enough time under these conditions to deplete CO, and so does not form significant quantities of \nthp. For final densities above $10^4 \pcc$, $t_{\rm dense} \sim t_{\rm freeze}$, and CO is sufficiently depleted for substantial quantities of \nthp{} to survive. As $t_{\rm freeze}$ scales as $\nh^{-1}$, even tracer particles with $t_{\rm dense} < 0.34 \myr$ can deplete most of their CO if their final density is above $10^4 \pcc$, as they only need to spend a small fraction of time at the higher density in order for freeze-out to proceed to completion.}

\subsection{Detectability of line emission}
\label{sec:radex}

\begin{figure}
  \centering
  \includegraphics[width=\columnwidth]{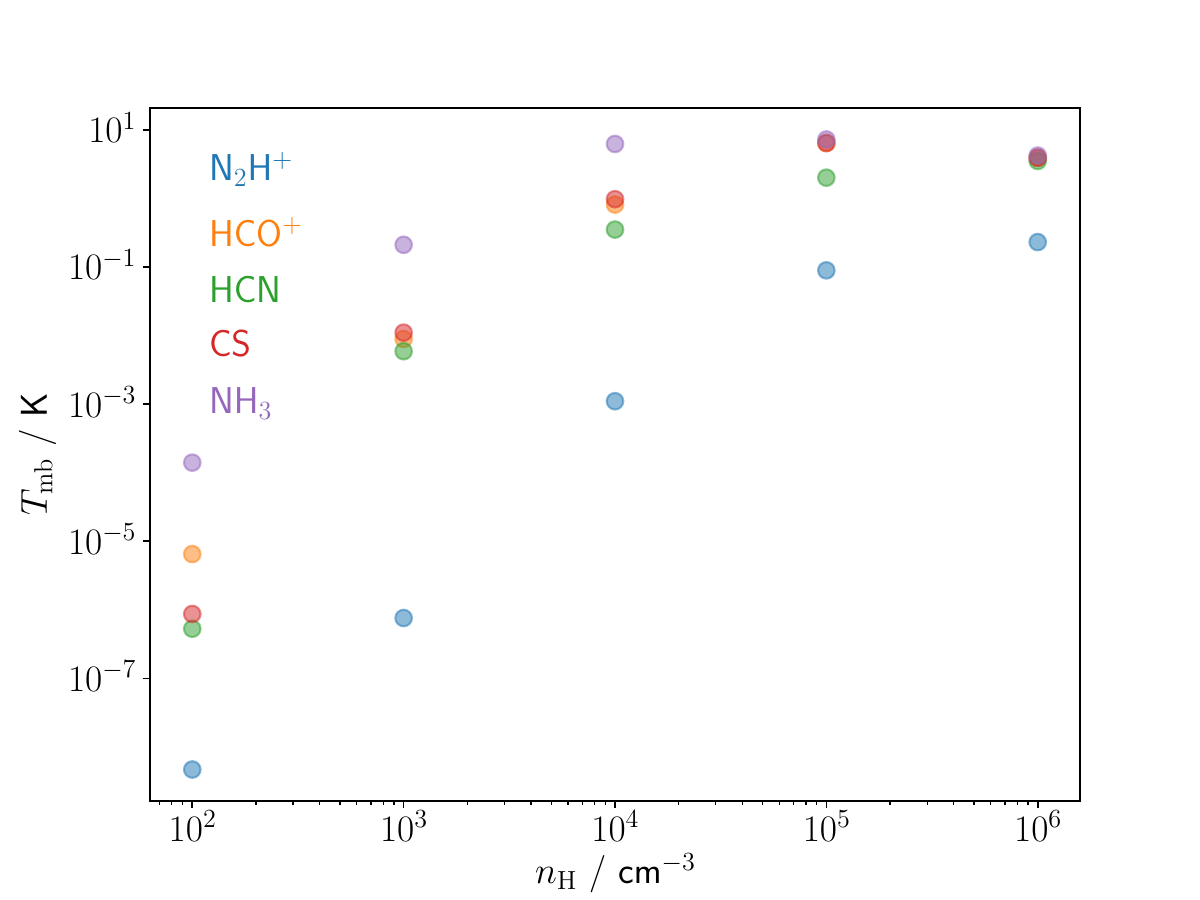}
  \caption{Main beam temperature versus density for the $J=1-0$ lines of \nthp{} (blue), HCO$^+$ (orange) and HCN (green), the $J=2-1$ line of CS (red), and the $(1,1)$ inversion transition of NH$_3$ (purple).}
  \label{fig:trad}
\end{figure}

\begin{table*}
  \centering
  \caption{Input parameters used for the radiative transfer modelling.}
  \begin{tabular}{ccccccccc}
    \hline
    $\nh$/$\pcc$ & $T$/$\kel$ & $\Nhcol$/$\pcs$ & $\sigma_v$/$\kms$ & $X($\nthp{}$)$ & $X($HCO$^+)$ & $X($HCN$)$ & $X($CS$)$ & $X($NH$_3)$ \\
    \hline
    $10^2$ & $47$ & $1.2 \times 10^{21}$ & $1.0$ & $10^{-14}$ & $10^{-12}$ & $10^{-12}$ & $10^{-12}$ & $10^{-10}$ \\
    $10^3$ & $23$ & $2.4 \times 10^{21}$ & $1.0$ &$10^{-13}$ & $10^{-10}$ & $10^{-9}$ & $10^{-9}$ & $10^{-8}$ \\
    $10^4$ & $15$ & $4.8 \times 10^{21}$ & $1.0$ &$10^{-11}$ & $10^{-9}$ & $10^{-8}$ & $10^{-8}$ & $10^{-7}$ \\
    $10^5$ & $10$ & $9.5 \times 10^{21}$ & $1.0$ &$10^{-10}$ & $10^{-8}$ & $10^{-8}$ & $10^{-7}$ & $10^{-6}$ \\
    $10^6$ & $7$ & $1.9 \times 10^{22}$ & $1.0$ &$10^{-10}$ & $10^{-8}$ & $10^{-8}$ & $10^{-7}$ & $10^{-6}$ \\
    \hline
  \end{tabular}
  \label{tab:rad}
\end{table*}

We estimate the intensity of molecular emission lines as a function of volume density using {\sc radex} \citep{vandertak2007}, {assuming a uniform slab geometry. The required input parameters are the volume density and temperature of the gas, the column density of the molecule in question, and the velocity dispersion in the slab. Both the gas temperature and the total column density are quite tightly correlated with volume density in our simulations, so we adopt the power-law fits to the average relationships between these quantities from \citet{priestley2023c} to obtain temperature and total column\footnote{Strictly speaking, this is the effective shielding column, not the line-of-sight value required by the radiative transfer model. However, the difference between these two quantities is typically small enough to make no qualitative difference to our results \citep{clark2014}.} as a function of density. We then convert the total column to the column density of the relevant molecule using its typical abundance in gas of the required volume density, estimated from Figures \ref{fig:peakmolv7} and \ref{fig:peakmolv7_2}. As there as no equivalent relationship between velocity dispersion and gas density, we assume a value of $1.0 \kms$ typical of molecular clouds, and note that observed line widths do not appear to vary strongly with density \citep{tafalla2021}.} Input parameters used in the modelling are given in Table \ref{tab:rad}.

Figure \ref{fig:trad} shows the main beam temperatures as a function of density for the $J=1-0$ lines of \nthp, HCO$^+$ and HCN, and the $J=2-1$ CS line, all of which are commonly observed in $3 \, {\rm mm}$ line surveys \citep[e.g.][]{pety2017,kauffmann2017,tafalla2021} and {assumed to originate from the higher-density regions of molecular clouds}. We also include the $(1,1)$ NH$_3$ transition, another line thought to trace higher-density gas than CO, and for which observational data is similarly plentiful \citep{ragan2011,ragan2012,friesen2017,feher2022}. The \nthp{} and HCN intensities correspond to the $(1,1,2)-(0,1,0)$ and $(1,0)-(0,1)$ hyperfine components, respectively; other hyperfine components {typically have intensities} within a factor of a few of each other, so the exact choice of transition does not affect our conclusions.

At a density of $100 \pcc$, all lines other than NH$_3$ are extremely weak ($<10^{-5} \kel$), but by $10^3 \pcc$ the HCN, HCO$^+$ and CS lines have strengthened to $\sim 10^{-2} \kel$ and NH$_3$ is above $0.1 \kel$. {These intensities are well within reach of modern telescopes with a modest\footnote{Using the IRAM $30 \, {\rm m}$ sensitivity calculator (https://oms.iram.fr/tse/), we estimate that a $3 \, {\rm mm}$ line reaching an intensity of a few $10^{-2} \kel$ would be detectable (exceeding the sensitivity limit by a factor of a few) in $\lesssim 10 \, {\rm hr}$ of telescope time.}} investment of observing time. \nthp{} remains several orders of magnitude fainter than this {at $10^3 \pcc$}, and does not even begin to approach comparable line brightnesses until a density of $10^4 \pcc$. As noted above, this corresponds almost exactly to the density at which material appears to become gravitationally bound. By a quirk of its chemistry, and unlike any other molecular species we have investigated, \nthp{} is only detectable in line emission in gas which will inevitably form stars.

\section{Discussion}

The fact that \nthp{} {specifically traces gas which is in the process of forming stars}, unlike most of the other species commonly used in this role, is due to its chemical behaviour. \nthp{} reacts {rapidly} with CO to form HCO$^+$ and N$_2$, and due to the high abundance of CO at even moderate densities, this is typically the main route of destruction. The \nthp{} abundance is therefore kept low until CO begins to deplete onto grain surfaces in significant quantities, which only begins to occur above {a density} of $10^4 \pcc$ (\citealt{tafalla2002,tafalla2004,caselli2002}; see also \citealt{priestley2023c}), {and requires that the gas remains at or above this density for a significant period of time (Section \ref{sec:freeze})}. No equivalent reaction with CO exists for the other species, meaning that they exist in detectable quantities below what appears to be the star formation threshold, at least in the simulations presented here.

One implication of our results is that although molecules such as HCN and HCO$^+$ do trace the material which will eventually go on to form stars, they also trace gas that is only transiently shocked to moderately high density and so plays no part in the star formation process. This makes them unreliable as tracers of star-forming gas. At first glance, this seems incompatible with the tight correlation between HCN luminosity and the star formation rate \citep{gao2004,neumann2023}, which has been interpreted as HCN directly tracing the gas involved with star formation. However, the HCN $J=1-0$ line intensity is known to be almost-linearly correlated with the total column of gas \citep{tafalla2021}, and therefore the mass. The correlation with star formation rate may therefore just be a correlation with the amount of molecular gas, rather than any specifically star-forming component of clouds. This would imply that the mass fraction of genuinely dense material is more-or-less constant, at least when averaged over large enough scales, which appears to be borne out by recent observational work \citep{garcia2023,jimenez2023}.

On smaller scales, when the structure of individual molecular clouds can be resolved, the distinction between material traced by HCN and by \nthp{} is likely to become important, as the ratio between their lines is clearly {\it not} constant \citep{pety2017,tafalla2021}. If HCN emission is assumed to trace {star-forming gas}, one might infer that the V14 model has a small but non-zero potential for star formation, despite the fact that it is actually in the process of dispersing. \citet{tafalla2021} find a near-linear correlation between the strength of the \nthp{} $J=1-0$ line and the total column density of material when $\Nhcol \gtrsim 3 \times 10^{22} \pcs$. We suggest that this can be used to estimate the mass of material corresponding to the immediate reservoir for star formation, $M_{\rm SF}$, via
\begin{equation}
  M_{\rm SF} = 310 \, \left( L_{\rm N_2H^+} / \kel \kms \pc^2 \right) \msun,
\end{equation}
where $L_{\rm N_2H^+}$ is the luminosity of the \nthp{} $J=1-0$ line integrated over all hyperfine components, and the proportionality constant is converted from the column density-line intensity scaling factor $X_{\rm N_2H^+} = 1.4 \times 10^{22} \pcs \, (\kel \kms)^{-1}$ found by \citet{tafalla2021}. If this material is collapsing on approximately the freefall time for gas at a density of $10^4 \pcc$, $t_{\rm ff} \sim 0.4 \myr$, the star formation rate can be estimated as $M_{\rm SF}/t_{\rm ff}$, giving
\begin{equation}
  {\rm SFR}_{\rm N_2H^+} = 775 \, \left( L_{\rm N_2H^+} / \kel \kms \pc^2 \right) \msun \myr^{-1}.
\end{equation}
Based on the delay between the formation of dense gas and sink particles in Figure \ref{fig:sfrate}, this quantity can be thought of as representing the expected star formation rate over the next $\sim \myr$ of evolution, rather than an estimate of previous activity as provided by methods such as protostar counts.

Although our simulations do not include protostellar feedback, heating by newly-formed protostars is expected to rapidly desorb molecules from grain surfaces in their vicinity \citep[e.g.][]{viti2004}, in particular CO. This increases the destruction rate of \nthp{}, and should eventually reduce its abundance back to undetectable levels. In reality, \nthp{} and even its deuterated isotopologues are commonly detected towards regions of active star formation \citep{fontani2011,friesen2013,cosentino2023}, so its presence is not uniquely associated with material that is {\it about} to form stars, but also with material that already has. We do not consider this distinction hugely important: \nthp{} still remains the best indicator of the potential for star formation of a molecular cloud, regardless of whether that potential has already been realised or not.

Finally, we note that our results are based on cloud properties, a radiation field and an ionisation rate appropriate for the Solar neighbourhood, and may not apply in more extreme environments. Unlike most other observational studies \citep{pety2017,kauffmann2017,tafalla2021}, \citet{barnes2020} find that \nthp{} behaves quite similarly to HCN in the W49 complex, a region of intense star formation activity. It is possible that an enhanced formation rate due to more highly ionised gas, or a reduced destruction rate due to greater photodissociation of CO, could change the chemical behaviour of \nthp{} to something closer to that of HCN, making it more readily detectable at moderate densities. However, it {also} seems possible that these same effects would change the threshold density for star formation, so it is unclear what impact this would have on the properties of various molecules as tracers. While this is beyond the scope of the present paper (the chemical network is only currently calibrated for Solar neighbourhood conditions), we plan to investigate this topic more extensively in future work.

\section{Conclusions}

We have post-processed hydrodynamical simulations of molecular clouds with a full time-dependent chemical network, in order to investigate the ability of various molecular species to trace {the reservoirs of material for star formation}. We find that many molecules commonly {used for this purpose}, such as HCN and HCO$^+$, exist in substantial quantities in gas which is presently at high density ($\nh \gtrsim 10^3 \pcc$), but only transiently, and which will eventually return to lower densities without undergoing any star formation. \nthp, by contrast, is only present in detectable amounts when the volume density of the gas exceeds $10^4 \pcc$, due to the onset of CO freeze-out. This coincides {closely} with the threshold density at which cloud material {typically} becomes gravitationally bound: the material traced by \nthp{} will inevitably form stars (if it is not doing so already), unless disrupted by some outside process. Regions of clouds detected observationally in \nthp{} therefore correspond directly to sites of current and future star formation, whereas other molecular tracers will give a misleading impression of a cloud's potential to form stars.

\section*{Acknowledgements}

FDP, PCC, SER and OF acknowledge the support of a consolidated grant (ST/K00926/1) from the UK Science and Technology Facilities Council (STFC). SCOG and RSK acknowledge funding from the European Research Council (ERC) via the ERC Synergy Grant “ECOGAL-Understanding our Galactic ecosystem: From the disk of the Milky Way to the formation sites of stars and planets” (project ID 855130, from the Heidelberg Cluster of Excellence (EXC 2181 - 390900948) “STRUCTURES: A unifying approach to emergent phenomena in the physical world, mathematics, and complex data”, funded by the German Excellence Strategy, and from the German Ministry for Economic Affairs and Climate Action in project ``MAINN'' (funding ID 50OO2206). The team in Heidelberg also thanks for computing resources provided by {\em The L\"{a}nd} through bwHPC and DFG through grant INST 35/1134-1 FUGG and for data storage at SDS@hd through grant INST 35/1314-1 FUGG. This research was undertaken using the supercomputing facilities at Cardiff University operated by Advanced Research Computing at Cardiff (ARCCA) on behalf of the Cardiff Supercomputing Facility and the Supercomputing Wales (SCW) project. We acknowledge the support of the latter, which is part-funded by the European Regional Development Fund (ERDF) via the Welsh Government.

\section*{Data Availability}
The data underlying this article will be shared on request.

\bibliographystyle{mnras}
\bibliography{densegas}


\bsp	
\label{lastpage}
\end{document}